# Proposing a conceptual framework: social media listening for public health behavior

Shu-Feng Tsao[1], Helen Chen[1], Samanthe Meyer[1], Zahid A. Butt[1*]

*1 School of Public Health Sciences, Faculty of Health, University of Waterloo, Waterloo, Ontario, Canada*

*\* Corresponding author*

## Abstract

Existing communications and behavioural theories have been adopted to address health infodemics. Although various theories and models have been used to investigate the COVID-19 pandemic, there is no framework specially designed for social listening or infodemiological studies using social media data and natural language processing techniques. This study aimed to propose a novel yet theory-based conceptual framework for infodemiological research. We collected theories and models used in COVID-19 related studies published in peer-reviewed journals. The theories and models ranged from health behaviours, communications, to infodemics. They are analysed and critiqued for their components, followed by proposing a conceptual framework with a demonstration. We reviewed Health Belief Model, Theory of Planned Behaviour/Reasoned Action, Communication for Behavioural Impact, Transtheoretical Model, Uses and Gratifications Theory, Social Judgment Theory, Risk Information Seeking and Processing Model, Behavioural and Social Drivers, and Hype Loop. Accordingly, we proposed our 'Social Media Listening for Public Health Behaviour' Conceptual Framework by not only integrating important attributes of existing theories, but also adding new attributes. The proposed conceptual framework was demonstrated in the Freedom Convoy social media listening. The proposed conceptual framework can be used to better understand public discourse on social media, and it can be integrated with other data analyses to gather a more comprehensive picture. The framework will continue to be revised and adopted as health infodemics evolve.

## Introduction

The World Health Organization (WHO) has consistently reiterated the widespread and multifaced nature of health infodemics and their harmful consequences throughout the pandemic (Wilhelm et al., 2022). The WHO initiated and hosted infodemic conferences and trainings since early 2020 to address increasingly complex health infodemics (Calleja et al., 2021; Tangcharoensathien et al., 2020; Wilhelm et al., 2022). The WHO's technical consultation has led to a framework to manage infodemics (Tangcharoensathien et al., 2020). Another framework that categorizes research agenda for infodemic management was developed from the first WHO's infodemic conference (Calleja et al., 2021). Before infodemics can be managed, it is necessary to measure and understand them. Over the course of the COVID-19 pandemic, recent systematic reviews have shown that health infodemics, especially health misinformation, have been prevalent and far-reaching on social media before and during the pandemic (Borges do Nascimento et al., 2022; Suarez-Lledo & Alvarez-Galvez, 2021; Wang et al., 2019). Depending on social media platforms, health misinformation can account for less than 1% to almost 30% of user-generated contents (Borges do Nascimento et al., 2022). Vaccine hesitancy fuelled by health misinformation has accounted for over 30% of the studies included in the systematic reviews (Suarez-Lledo & Alvarez-Galvez, 2021; Wang et al., 2019). However, given researchers from diverse backgrounds with different expertise, it is unsurprising that various theories have been used to guide studies of health infodemics (Ngai et al., 2015). Different theories have suggested inconclusive predictors, mediators, and moderators, but scholars have constantly regarded behavioural intentions or behaviours as ultimate outcomes, yet their measurements have varied (Ngai et al., 2015). Additionally, further research is needed to understand how online infodemics have influenced offline behavioural intentions or behaviours (Calleja et al., 2021). The WHO has repeatedly called for multidisciplinary collaborations since professionals in communications, neuroscience, and digital marketing have long studied how social media have manipulated people's behaviours (Aral, 2020; Wilhelm et al., 2022).

With the advancement in natural language processing (NLP), infodemiological research applying different NLP techniques to analyze social media data to understand public discourse—called social listening—has exponentiated. For example, the WHO has developed and deployed a "Early AI-supported Response with Social Listening" (EARS) platform to identify emerging information voids following WHO's terminologies (Purnat, Vacca, et al., 2021; Purnat, Wilson, et al., 2021). Nonetheless, existing social listening tools, given their marketing-driven designs, need great customizations to meet the needs for infodemic social listening like the EARS platform (Purnat, Vacca, et al., 2021; Purnat, Wilson, et al., 2021). In a public health crisis, health professionals need a tool that can efficiently harness and analyse tremendous amounts of online data to understand the public discussions in timely manners since qualitative analysis is time-consuming. Latest NLP techniques, including but not limited to topic modelling, sentiment analysis, and stance detection, have been used in infodemic social listening (ALDayel & Magdy, 2021; Medhat et al., 2014; Vayansky & Kumar, 2020). Although improvements are still needed to decrease misclassifications in these supervised and unsupervised NLP techniques, their accuracies have been acceptable so far. These NLP techniques are commonly used as a screening layer to quickly understand public discourse at a superficial level, followed by qualitative analysis to make sense, enhance understanding, or identify information voids from the conversations. Such integrated social listening, on average, can be done on a weekly basis, along with other data sources (Boender et al., 2022).

It is understandable that, in the beginning of the COVID-19 pandemic and infodemic, researchers agreed to adapting existing health theories, such as the health belief model (HBM) and social cognitive theory (SCT), and social-ecological model (SEM), and tools to overcome challenges in generating new tools given limited resources (Calleja et al., 2021; Tangcharoensathien et al., 2020). Although these health theories have been long established, most of them are developed before the existence of social media (Aral, 2020). Ubiquitous social media has changed how people consume and behave upon online health information for better or worse (Aral, 2020). Dr. Schillinger et al.'s Social media and Public Health Epidemic and Response (SPHERE) model (2020) and Dr. Aral's Hype Loop (2020) have demonstrated that social media have both perils and merits. That is, social media can help people make informed decisions while spreading harmful misleading information (Aral, 2022; Schillinger et al., 2020). The WHO has recommended that social listening for infodemic management should be incorporated into future pandemic preparedness (Tangcharoensathien, et al., 2020; Wilhelm et al., 2022).

During the pandemic, social listening has mostly been reactive than proactive. Health professionals and public health organizations were rushed to debunk misinformation while competing for people's attention to urge people to follow evidence-based preventive behaviours during uncertainties (Nan et al., 2022; Vraga & Jacobsen, 2020). Although many lessons have been learned regarding health infodemics using existing theories and tools, there is a need to carry out social listening in a systematic way based on a novel theoretical framework for health researchers. Except Dr. Aral's Hype Loop (2020), there are limitations in current theories or frameworks developed before the existence of social media. Therefore, the objective of this paper was to propose a conceptual framework that helps monitor public discourse on social media and behaviours for future infodemiological research and possible utility of the proposed conceptual framework.

## Methods

Borsboom, et al.'s (2021) theory construction methodology (TCM) was adapted to help develop a conceptual framework. According to TCM, there are five steps: (1) identification of relevant phenomena, (2) development of a proto theory, (3) formation of a formal model, (4) adequacy evaluation of the formal model, and (5) assessment of overall worth of the formal model. Firstly, we identified health infodemics on social media as a phenomenon of interest since we were especially interested in how online information on social media has influenced people's behavioral intentions or behaviors in a public health emergency. Next, we conducted a theory synthesis (Walker & Avant, 2018) to develop a conceptual framework as the TCM's second and third steps were combined. We searched PubMed, Scopus, PsycINFO, and Google Scholar for theories used in reviews and original research papers written in English published in peer-reviewed journals from December 2020 to December 2022. Keywords included 'social media,' 'online discussion,' 'public discourse,' 'behavior,' 'intention,' 'attitude,' 'perception,' 'theory,' 'model,' 'framework' and their synonyms, but explicitly excluded 'conspiracy theory' in the search. Reviews were prioritized for extractions and reading because certain theories have been commonly used in the COVID-19 related studies in health behavioral science, communications, and infodemic management. We included theories with outcomes as health behavioral intentions or behaviors at individual level and beyond. The search for relevant theories in this process was non-exhaustive, but the results were representative of the health infodemic research conducted thus far. A total of 13 theories are included for Walker and Avant's theory synthesis (2018). After the conceptual framework was formulated, a demonstration was conducted to check and evaluate the overall conceptual framework to meet the last two steps in the TCM.

# Results

## *Synthesis of Theories*

Table 1 shows the thirteen theories included in this study. As expected, the health belief model (HBM) has been widely employed since one systematic review reported that HBM was used in 126 quantitative studies about the COVID-19 Vaccine Hesitancy over two years (Limbu et al., 2022). It is also expected that some existing theories were combined or adopted by researchers to investigate complex and multifaceted health infodemics in various studies. For example, the theory of planned behaviour (TPB) itself is an extension of the theory of reasoned action (TRA) (U S Dept of Health and Human Services, 2018). Additionally, TPB was combined with the heuristic systematic model (HSM) to create the risk information seeking and processing model (RISP) (Yang, Aloe, et al., 2014; Yang, Liu, et al., 2022), or integrated with the uses and gratifications theory to investigate information-sharing behaviours (Malik et al., 2021). Furthermore, Scannell et al. (2021) weaved the social judgement theory, elaboration likelihood model of persuasion (ELM), and extended parallel process model (EPPM) to understand how persuasive COVID-19 vaccine (mis)information was to convince people, implicitly affecting their behaviours (Scannell et al., 2021). Overall, it has demonstrated that a theoretical approach may no longer be sufficient to address the complexity of health infodemics.

| Theory/Model | Focus | Constructs |
|---|---|---|
| Behavioral and Social Drivers | Behavior | Confidence, Motivation, and Behavior |
| Capability, Opportunity and Motivation lead to Behavior | Behavior | Capability, Opportunity, Motivation, and Behavior |
| Elaboration Likelihood Model | Attitude or Behavior | Motivation, Ability, and Opportunity to decide Central route or Peripheral route |
| Extended Parallel Process Model | Behavior | Threat and Efficacy |
| Health Belief Model | Behavior | Perceived susceptibility, Perceived severity, Perceived benefits, Perceived barriers, Modifying variables, Cues to action, and Self-efficacy |
| Risk Information Seeking and Processing Model | Attitude or Behavior | Combine both theory of planned behaviour and heuristic systematic model |
| Social Cognitive Theory | Behavior | Behavioral capability, Observational Learning, Reinforcements, Expectations, Self-efficacy, and Reciprocal Determinism |
| Social Judgment Theory | Attitude | Latitude of Acceptance, Latitude of Non-commitment, and Latitude of Rejection |
| The Hype Loop | Behavior | Consume, Act, Sense, and Suggest |

| Theory of Planned Behavior | Behavior | Attitudes, Subjective norm, Perceived behavioral control, Behavioral intention, and Behavior |
| Theory of Reasoned Action | Behavior | Attitudes, Subjective norm, Behavioral intention, and Behavior |
| Transtheoretical Model | Behavior | Precontemplation, Contemplation, Preparation, Action, Maintenance, and Termination |
| Uses and Gratifications Theory | Behavior | Cognitive need, Affective need, Personal integrative need, Social integrative, and Tension release need |

Table 1: Theories and models used in health infodemic research in the context of the COVID-19 pandemic.

Of these theories, several factors across theories have repeatedly been shown to affect the outcome (i.e., behavior). Although they are described in different terms, they can be used interchangeably in most contexts. For instance, the "self-efficacy" in HBM and social cognitive theory (SCT) has shared a similar meaning with "confidence" in the behavioral and social drivers (BeSD) of vaccination, "perceived behavioral control" in TPB, and "efficacy" in EPPM. If the meaning is extended further, it can also represent "capability" in the model of capability, opportunity, and motivation lead to behavior (COM-B), "ability" in ELM, "behavioral capability" in SCT, "Act" in the Hyper Loop, and "behavioral intention" in TPB/TRA, and the Transtheoretical Model. Another group of terms—altitude, perceptions, and motivation—can also share comparable meanings, although they have different definitions in a dictionary. Five of the thirteen theories include "attitude," another three theories consist of "motivation," and the other two theories involve perceived variables that are associated with the outcome. In general, these words have suggested people's views in consistent or in contrast to given health information. These terms have also suggested that there are gaps between "self-efficacy' and "(cap)ability," "perception" and "reality," or "subjectivity" and "objectivity." However, it can be challenging to distinguish them because they shape each other. That is, I believe I can do it this time (i.e., subjective self-efficacy or perception) because I did it before (i.e., an objective real action). Now I get it done (i.e., objective real action), so I know I will be able to do it next time (i.e., subjective self-efficacy or perception), with or without extra preparation or practice. It becomes greatly interrelated and thus these two may no longer be discernible, or it is too difficult to measure them separately. Similarly, attitudes and perceptions may be indistinguishable as they both imply motivations or intentions for behavioral uptake or changes.

Although almost all theories focus on individual behaviors, factors beyond individuals are also important to be considered and yet these social determinant factors can be difficult to measure or imprecise based on self-reported measurements (Beauchamp, et al., 2018; Carpenter, 2010; Gagne & Godin, 2000; J. Kitchen, et al., 2014; Marks, 2020; Nigg, et al., 2011; Schunk & DiBenedetto, 2020). However, existing models, such as HBM, SCT, and BeSD, can incorporate variables beyond personal levels to infer the outcome. Nonetheless, unlike EPPM, these behavioral models don't explicitly measure emotional variables, although they might be inferred in variables related to self-efficacy, perceptions, or subjective norms. One of implicit assumptions in these theories is that people can logically determine and behave to mitigate risks if they perceive greater threats or susceptibility to themselves. According to latest infodemic and social media research (Aral, 2020; Azer, et al., 2021; Purnat, et al., 2021), unfortunately, behaviors may not be completely driven by rational reasoning; otherwise, panic buying during the COVID-19 pandemic is not supposed to happen (Naeem, 2021). Prior studies have evidently shown how social media, given their artificial algorithm designs, can manipulate or help spread emotional posts, making it contagious at large (Aral, 2020; Oh, et al., 2021; Steinert, 2020). Therefore, emotion should also be taken into account when inferring behaviors, similar to perception, attitude, motivations, and others.

Given limitations and gaps identified in existing theories and frameworks, a new framework is needed to reflect the current complex infodemic issues in today's information ecosystem ( Calleja et al., 2021; Tangcharoensathien et al., 2020; Wilhelm et al., 2022). The new conceptual framework should incorporate theories from the communication field because it will improve health professionals' understanding of public discourses. In addition, attributes measuring attitudes and emotions are included in the proposed conceptual framework: Social Media Infodemic Listening (SoMeIL) for Public Health Behavior.

## Proposed Conceptual Framework

We propose a novel conceptual framework—SoMeIL for public health behavior (Figure 3-1)—to address these issues. Our framework aims to investigate how people's emotions and attitudes are associated with their online behaviors on social media, and eventually their offline behaviors in the real world. In other words, our proposed framework can help researchers to better understand the public discourse and to better infer collective behavioral intentions or behaviors. Double arrows illustrate potential associations these five constructs have with each other. Blurry boundaries and faded colors demonstrate that the components can happen both online and offline simultaneously. Unlike existing theories, our framework no longer assumes rational judgments and behaviors. In the following sections, we will introduce and explain each construct illustrated in our proposed conceptual framework, along with some limitations in social media data or NLP techniques when researchers use them.

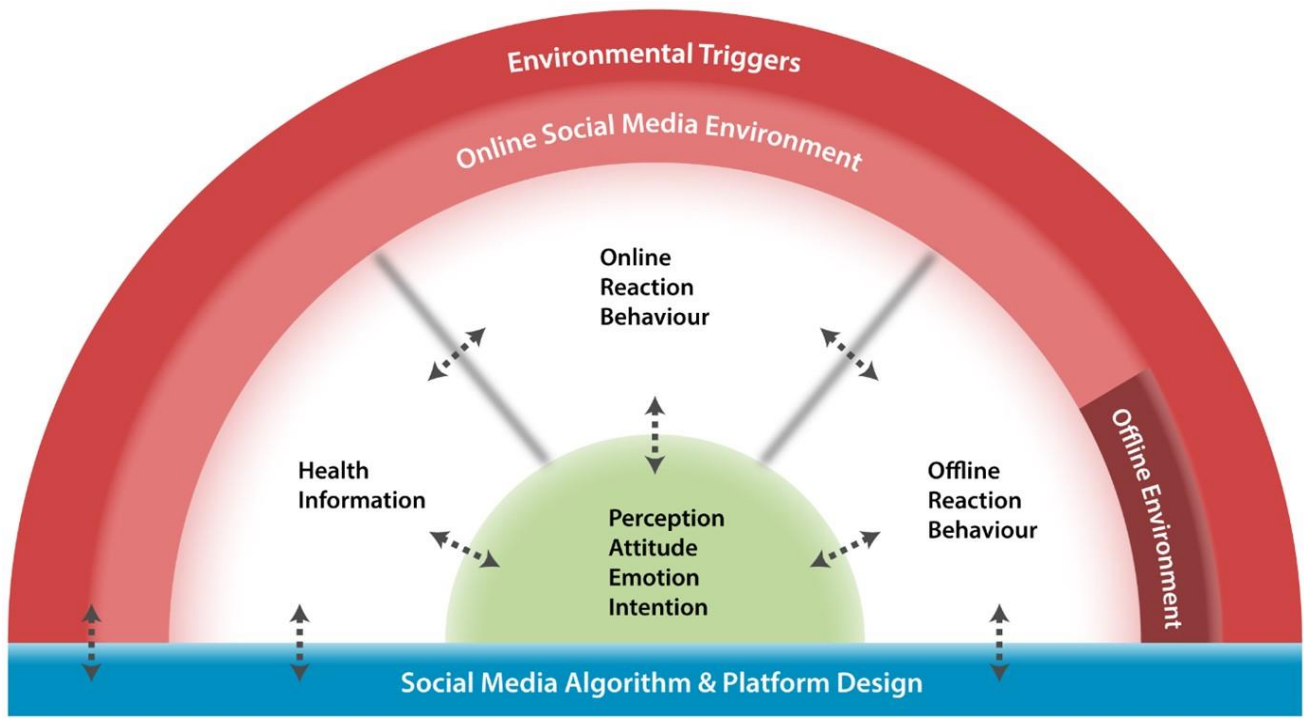

Figure 1: Social media infodemic listening (SoMeIL) for public health behavior conceptual framework

As Dr. Aral (2020) has demonstrated in his research, the social media algorithm is input with user attributes (Table 2), such as the demographics and historical behavioural data to connect friends or 'recommended' posts to users based on similarity instead of heterogeneity (Aral, 2020). Studies have evidently shown that social media algorithms are intentionally designed to be addictive and affective (Aral, 2020; Azer, et al., 2021). The issue is further compounded by highly personalized user experiences on social media given people's digital footprints, encouraging echo chambers or polarization (Aral, 2020). Coupled with its engagement design, such as 'like' and 'follow' buttons, social media have kept their users spending more time on the platform as 'engagements' (Aral, 2020). Such characteristics is defined as 'user attributes on social media algorithms' in the proposed conceptual framework (Figure 1). However, since user attributes are voluntarily input by users when they first create their accounts, most attributes (Table 2) are optional, and values can be fictitious. In other words, they all can be missing data, or even untrue when values are not missing, although correct values exist. Some social media platforms require users to enter only their email and password to create an account with a username without any other details. Therefore, the users can remain primarily anonymous or unverified on the platform. Geolocations is a special issue for researchers when modelling disease outbreaks or heat maps using Twitter data (Dredze et al., 2013; Sloan & Morgan, 2015; Stock, 2018). For example, tweets tagged with explicit geolocation can vary from less than 1% to approximately 4% of data collected from Twitter (Dredze et al., 2013; Sloan & Morgan, 2015), depending on data collection methods and the amount of data collected. Although there are many machine learning (ML) techniques to infer geolocations for Twitter data (Dredze et al., Stock, 2018), they are not as precise or comparable as Internet Protocol (IP) addresses. Futhermore, public discussions related to vaccinations on social media have become more polarized over time, for instance (Jiang et al., 2021; Schmidt et al., 2018; Yuan et al., 2019; Mønsted & Lehmann, 2022; Rathje et al., 2022). Studies have demonstrated that user attributes, such as users' political party affiliations, religious affiliations, and who they follow (i.e., following), can potentially indicate their ideologies or attitudes toward vaccinations (Jiang et al., 2021; Mønsted & Lehmann, 2022; Rathje et al., 2022; Schmidt et al., 2018; Yuan et al., 2019). Similar to the geolocation issues, researchers may not have direct access to collect these attributes. If users enter some information within these attributes, their accuracies remain uncertain. Additionally, even if researchers apply advanced ML techniques to infer these attributes, these techniques may be unable to generalize to other studies or social media platforms with different users' characteristics (Shakeri Hossein Abad et al., 2022).

| Components | Attributes |
|---|---|
| User attribute on social media | - Age<br>- Sex<br>- Geolocation<br>- Income<br>- Education<br>- Occupation<br>- Party affiliations<br>- Region affiliations<br>- Following<br>- Followed |
| Inferred Intention | - Attitude<br>  - Acceptance<br>  - Non-commitment<br>  - Rejection<br>- Emotion<br>  - Positive<br>  - Negative<br>  - Neutral<br>  - Mixed |

|  | • Perception |
|  | • Ideology |
| Online Reaction Behavior | • Share |
|  | • Like/dislike |
|  | • Comment |
|  | • Post |
|  | • Bookmark |
|  | • Nothing |
| Offline Reaction Behavior | • Agreement |
|  | • Disagreement |

Table 2: Attributes

Next, we define 'online behaviour' as it occurs 'after' a user views a social media post. We can measure collective online behaviours via the numbers of likes, shares, and others (Table 2). These attributes are not mutually exclusive because a person can have multiple reactions after viewing a post. Besides, we add an attribute called "nothing" to reflect that an individual may have no reaction at all, or a reaction that is not captured by the social media platform. For example, the user may laugh so hard in reality but doesn't even 'like' the post after viewing a hilarious post. The "nothing" attribute is theoretically the same as 'non respondent bias' in survey research. Although there are other digital tracking tools to help infer viewers without any online reactions, researchers have been unable to directly access or retrieve such information since social media companies can decide what information can be available to researchers. We are especially interested in online behaviour, or its propagation patterns because it can be used to infer or confirm collective inferred intentions, as measured by emotions, attitudes, or perceptions. For instance, digital marketing research on Twitter has long estimated the number of users sharing similar opinions (i.e., acceptance) by the number of likes and retweets of a given tweet, whereas disagreements (i.e., rejection) can be reflected by the number of replies (Salamander, 2022). Dividing the latter by the former, if the resulting 'Twitter ratio' is at least 0.5, it indicates positive or neutral responses, whereas below 0.5 suggests negative responses (Salamander, 2022). Therefore, by collecting and analysing the attributes within the online behaviours, scholars can better understand or estimate what inferred intentions of the 'quiet majority' users are since approximately 10% of users produce 90% of content on Twitter, for example (Silverthorne, 2009). Online behaviour can be used to infer people's behavioural intentions. For example, if someone tweeted that they would get COVID-19 vaccinated as soon as they became eligible, and the tweet resulted in 1,500 likes and 2,500 retweets, it was estimated approximately 3,501 pro-vaccine people. Nonetheless, the number can be an overestimation considering that reactions are not mutually exclusive, or an underestimate since Twitter users do not really represent a general population in a given region. Additionally, such estimations may not apply to other social media.

As explained in the theory synthesis, existing models have theorized that behaviours can be attributed to attitudes, perceptions, and emotions, but it has remained challenging to clearly distinguish them because they are interrelated and cannot be easily measured. Researchers have inferred associations among attitudes, perceptions, and emotions in various ways (ALDayel & Magdy, 2021; Bahamonde-Birke et al., 2015), but we decide to group them together in our framework as "inferred intention" In our opinion, it is unnecessary to distinguish them since they can be used interchangeably or along with each other in different contexts. It becomes more important to infer potential behavioural intentions using attitudes, perceptions, emotions, or ideologies. we have adopted SJT to infer intentions (Table 2) because this makes it more feasible when using NLP techniques to analyse social media data, especially in infodemiological studies. For example, when investigating public intentions toward COVID-19 vaccination, acceptance can be theoretically associated with pro-vaccine individuals, rejection probably suggests anti-vaccine people, and non-commitment might be regarded as a proxy for vaccine-hesitant people as evidenced by prior research (Nyawa, et al., 2022). Yet we acknowledge that there are limitations in this assumption, so we need to be careful in how we interpret data and ascribe intentions based on our categorization of individuals. To better understand public discourse on social media, a promising ML technique—stance detection—can be applied to infer whether or not people's attitudes toward a give topic (ALDayel & Magdy, 2021; Cao et al., 2022; Nyawa, et al., 2022). For example, whether or not people support or oppose the COVID-19 vaccination. In addition to stance detection (Cao et al., 2022), a common way to infer attitudes in existing infodemic studies involves topic modelling and sentiment analyses (ALDayel & Magdy, 2021; Medhat et al., 2014; Vayansky & Kumar, 2020). Depending on models of sentiment analyses, emotions can be categorized at basic levels (i.e., positive, neutral, and negative) or more detailed levels (e.g., sad, anger, happy, joy, etc.) (Nandwani & Verma, 2021; Zadra & Clore, 2011). However, according to our research experiences and other infodemic studies, sentiment analysis can still result in misclassifications regardless of levels (Davis & O'Flaherty, 2012; He, et al., 2022; Lee, et al., 2022). Therefore, our framework remains conventional to maintain emotions at basic levels with an additional level called "mixed" sentiment. The "mixed" attribute is added to address possible misclassifications in the "neutral" category resulting from sentiment analysis. When a tweet includes an approximately equivalent number of positive and negative words, it's classified as 'neutral' by the sentiment analysis. However, this doesn't mean the tweet is really 'neutral' because it can actually be 'positive,' 'negative,' or 'mixed' overall, depending on its context (Davis & O'Flaherty, 2012; He, et al., 2022; Lee, et al., 2022). Misclassifications often occur in ironic or humorous tweets (Davis & O'Flaherty, 2012; He, et al., 2022; Lee, et al., 2022). The 'mixed' feeling in our framework refers to an equal amount of positive and negative feelings expressed simultaneously in a tweet without being 'positive' or 'negative' overall. For instance, if someone tweets equal number of concerns and favours towards COVID-19 vaccines without explicit conclusions, this tweet can be regarded as 'mixed' by humans, but it's likely classified as 'neural' by sentiment analysis. Although we incorporate the 'mixed' attribute in our framework, we acknowledge that existing sentiment analyses have not been sophisticated or advanced enough to categorize such 'mixed' feelings. In addition, even humans cannot interpret 'mixed' feelings consistently given external social-cultural factors, similar to humours are different in different cultures. Therefore, improvements are still needed.

For the 'offline behaviour' shown in Figure 1, although boundaries between our physical and digital worlds have become less distinguishable, it remains unclear whether or not people really react upon information received from social media. Some may have consistent online and offline reaction behaviours, another may have contradictory online and offline reaction behaviors, and others may only have either online or offline reaction behaviours. Even if individuals tweet or like a tweet indicating that they are willing to get vaccinated, it remains inconclusive unless they later share a selfie or their vaccination record on social media to prove that they, in fact, get COVID-19 vaccinated. In this case, their offline behaviour matches their online reaction behaviour. Their offline behaviour is also adherent to public health interventions. Therefore, one's offline behaviour can be inferred in two ways: one is whether an individual's online and offline behaviours are consistent, and the other is whether their offline behaviour follows the public health interventions.

Our life has become more digitalized, and younger generations have tended to share their life on social media, but it has not made it easier to associate offline behaviour with online information due to privacy concerns and available social media data. Therefore, additional data resources or proxy measures will be needed to investigate associations between online information on social media and offline behaviours at population levels. For instance, administrative or census data regarding COVID-19 vaccine administration can be linked with social media data to estimate or confirm vaccination coverage in a region. Similar approaches can be applied when there are outbreaks of infectious diseases.

*Framework Evaluation*

According to TCM (Borsboom, et al., 2021), it is imperative to check the explanatory adequacy and evaluate overall worth of the proposed conceptual framework. Therefore, we demonstrated the application of the proposed conceptual framework using an unsupervised LDA topic modeling and qualitative thematic analysis. Table 3 shows the LDA-generated topics, and findings of the thematic analysis are summarized in Figure 2. The themes and sub-themes identified from the thematic analysis were not exactly aligned with individual topics resulting from the LDA topic modeling, but they have helped interpret the contexts of each topic. Each topic included tweets both supporting and opposing the Canadian Freedom Convoy protest with corresponding contexts. The first topic was primarily about people's attitudes toward the convoy, both agreeing or disagreeing with protestors given their political tendances or perceptions of individual freedom. The second topic showed people blamed politicians or political parties that did not align with their worldviews (i.e., pro- or against the convoy), with the Canadian prime minister being especially called out. The third topic was discourse related to the vaccine mandates imposed on truckers since the convoy supporters wanted to end the mandates that, they believed, violated their freedom right. The fourth topic was about the Ottawa police activities containing the protest that has blocked the city for a prolonged period of time. However, some against the rally criticized the lacking or weak Ottawa police enforcement in the first place. The last topic focused on fundraising efforts from the convoy supporters, such as the GoFundMe page.

| Topic | Top 15 Words |
| --- | --- |
| 1 | support, covid, stand, cdnpoli, govern, arrest, driver, news, call, peopl, rally, terrorist, tyranni, time, today |
| 2 | trudeau, like, world, video, thank, peac, honkhonk, love, look, speak, movement, share, power, lie, flag |
| 3 | mandat, protest, report, live, end, start, vaccin_mandate, want, stop, govern, vaccin, country, way, american, ottawa |
| 4 | ottawa, polic, day, come, break, ontario, weekend, week, protestor, head, kid, help, citi, thousand, actual |
| 5 | peopl, right, medium, know, need, donat, think, thing, gofundm, go, want, let, fund, watch, organ |

Table **Error! No text of specified style in document.**-1 Latent Dirichlet allocation topic modeling results

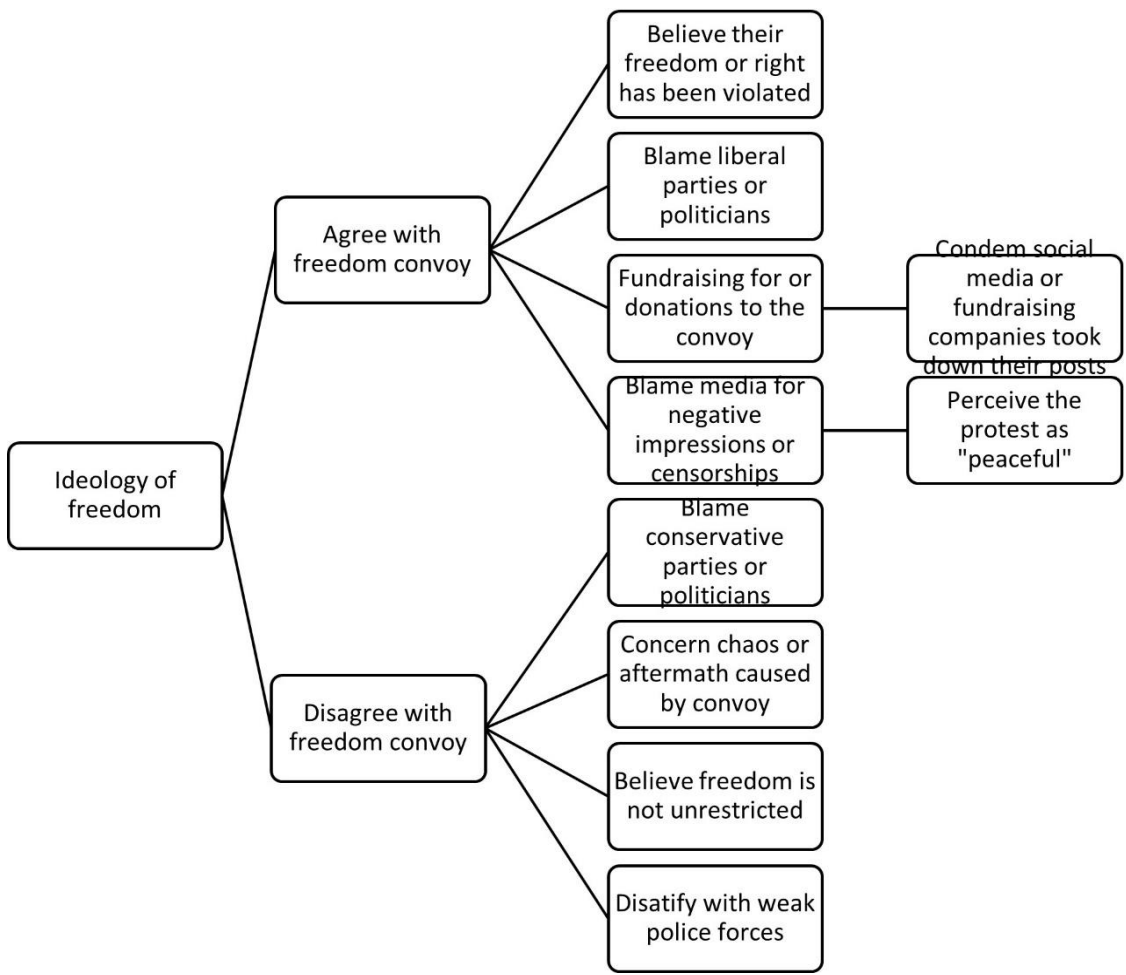

Figure2: Thematic analysis results

## Discussion

We proposed a new conceptual framework (Figure 1) consisting of five major constructs inspired from existing theories. Dashed boundaries indicate that online and offline environments have become less distinctive as information flows. Arrows represent potential associations among these components and how they influence or self-feed each other as the framework gives a sense of loop. Attributes of each construct can be inferred or measured via advanced NLP or ML techniques if data are available and of high quality. Although we have used NLP techniques to explain our conceptual framework throughout this paper based on our studies published during the COVID-19 pandemic (Huang et al., 2022; Tsao et al., 2022), the proposed framework is not limited to quantitative infodemiological research only. That is, the proposed conceptual framework can be applied in qualitative research.

In the example of Canadian Freedom Convoy, the results of our LDA topic modelling confirmed a previous study (Huang et al., 2022), the ideology of freedom has emerged as an overarching theme resulted from the qualitative thematic analysis, illustrating an explicit difference in freedom perceptions between the convoy supporters and opponents. This also appeared in each topic resulting from the LDA topic modelling. Among the convoy advocates, they strongly believed the vaccine mandates has fundamentally violated their constitutional right of freedom, as one tweet illustrated: '…the Convoy stands for freedom, and a democracy, as opposed to what we are experiencing…. Borderline China! No more mandates!' Similarly, some tweets articulated that they were not "anti-vaxxers" as news media labelled them. They said that they opposed "mandates" rather than against "vaccines," and the whole point, the protest supporters argued, was that they should have freedom to choose or decide for themselves rather than being forced to follow whatever the government had them do, such as the "zero COVID-19 policy" imposed by China on its own people. In contrast, people against the convoy implied such protesters selfish and believed that the freedom was not unrestricted, as one tweet explained: 'The thing is, the Charter has been tested over things like public health measures and workplace safety before, and it's not a right to do what you want, all the time, anywhere, with anyone.'

Given different viewpoints, it is not uncommon that each side criticized politicians or political parties at the other side since the ideology of freedom is also shaped by political affiliations. The protest proponents vilified the liberal politicians or parties, especially pointing to the Canada prime minister, as the following tweet demonstrated: 'When you allow a sick leader's ego to hold a country hostage you become a fascist sympathizer. My mother lived under Nazis occupation. She said the righteous ones are always the real fascists.' The same logic applied to people against the convoy as they blamed the conservative politicians or parties that supported the protest: 'Shameful for the Conservatives to embrace these loons. And not surprising.' Furthermore, people who supported the convoy criticized news media for negative reports or coverages by, for example, arguing: 'The media coverage of the #FreedomConvoy2022 has largely been nothing short of disgraceful disinformation. Having failed to ignore the convoy into oblivion, @XXX + the leftist press seized upon a dishonest and cliche narrative and set out to tell the story accordingly.'

In other words, the convoy supporters defended their protest not as terrible as the media painted them, and the media used political propaganda against them by imposing disproportionally negative biases. Some went even further to tweet the convoy was just 'peaceful' protesters as opposed to 'violent' lawbreakers, 'fringe minority,' 'far-right extremists,' or 'white supremacists' as the media called them. Given the perception of 'peaceful' protest, people backing the convoy also condemned the state of emergency declared by the Ottawa Mayor to use increasing police forces to break down the protest. Compared to the 'peaceful' convoy perceived by the supporters, the convoy opponents, consistent with the media, considered the protest 'violent' and caused numerous harmful chaos or aftermaths. For instance, one tweet showed its disapproval: 'So Canada's 'Freedom Convoy', opposing vax mandate for truckers, - Harrassed homeless shelter soup kitchen demanding food & assaulted a homeless person, - Defaced the Terry Fox Memorial Statue by draping it with an upside down CA - Stood on The Tomb of The Unknown Soldier WTAF?!'

People against the convoy also criticized the police. However, unlike criticisms from the convoy supporters, people who disagree with the protest perceived insufficient police enforcement to contain the protest before the state of emergency was declared. Although these people acknowledged that the truckers had the right to protest, the protesters should not just block Ottawa the way they wanted and disrupted residents' daily lives.

In addition, they condemned the police who showed supports for the convoy by not seriously enforcing the laws or donating to the convoy, just similar to the convoy proponents tweeted their supports via donations: 'In for $20, wish i could kick in a lot more...a patriot donated 10k anonymously, God Bless you!' In the meantime, there were fundraising tweets calling for donations. When the fundraising post was suspended and donations were frozen, the convoy supporters unleashed their anger and tried to find other alternatives. They heavily criticized social media that censored their posts, in addition to the fundraising websites that took down their pages. On the contrary, the convoy opponents showed support for donation suspensions.

The example has showed that using both the NLP topic modelling and qualitative thematic analysis, researchers could better understand public discourse given its contexts on a social media in a relatively shorter period compared to only thematic analysis or other qualitative studies. Furthermore, the demonstration has shown how to apply the conceptual framework to better understanding, or 'social media listening,' of certain events and to infer behavioural intentions more efficiently. The convoy supporters could be assumed that they had been relatively resistant to the COVID-19 vaccinations than the convoy opponents. In other words, the protesters were less likely to get vaccinated than others as evidence by their perceptions and ideology of freedom.

There are several limitations in our conceptual framework. Firstly, more evaluations need to be conducted since this is a new proposed conceptual framework. Although we showed the application of the proposed conceptual framework using the Canadian Freedom Convoy event, only English tweets were included. Therefore, the findings could not be generalized to other social media platforms since characteristics of the Twitter users may not be representative for others. The LDA topic modelling could not analyse non-textual data, including emojis, memes, and videos. This could result in misclassifications, although we mitigated this issue in the thematic analysis by randomly sampling analysing the raw tweets. Furthermore, given that the proposed framework primarily focuses on social media, it is acknowledged that this proposed framework can only be useful in more digitalized populations, cultures, or nations. Besides, with new social media platforms popping up, data formats and types can change given different platform designs. Therefore, the proposed framework may need to be revised to reflect and investigate non-textual data, such as videos and images. Although there are advanced NLP and ML techniques that can analyse videos and images, they have not been well adapted in current infodemiological or social listening studies. Lastly, each social media has different user characteristics. Therefore, social media data can be biased. Researchers will need to be careful when interpreting findings from different social media platforms even with our proposed conceptual framework.

Although existing health behaviours, communications, and latest infodemic theories have been used in infodemiological studies, these theories have not reflected well the distinctive nature of social media in the current complex information ecosystems. Therefore, we proposed a novel conceptual framework—social media infodemic listening (SoMeIL) for public health behaviour—to help future infodemiological research. We acknowledge that our conceptual framework still needs validations for its efficacy, safety and usability. We anticipate the proposed framework will be revised as more studies will be conducted in the future.